\documentclass[10pt,final,journal,twocolumn]{IEEEtran}
\usepackage{amsmath}
\usepackage{subfigure}
\usepackage{graphicx}
\usepackage{latexsym}
\usepackage{amssymb,amsbsy,verbatim,array}
\usepackage{epstopdf}
\usepackage{cite}
\usepackage{color}
\usepackage{cleveref}
\usepackage{times} 
\usepackage[top=0.7in, left=1in, right=1in, bottom=1in]{geometry}
\newtheorem{remark}{Remark}
\newtheorem{theorem}{Theorem}

\everymath{\displaystyle}
\begin{document}
\title{FAS-assisted Wireless Powered Communication Systems}

\author{Xiazhi Lai,  Kangda Zhi, Wanyi Li, Tuo Wu, Cunhua Pan, \emph{Senior Member, IEEE},  and\\ Maged Elkashlan , \emph{Senior Member, IEEE}
\thanks{\emph{(Corresponding author: Tuo Wu and Cunhua Pan.)}}
\thanks{X. Lai and W. Li are with the School of Computer Science, Guangdong University of Education, Guangzhou, Guangdong, China (E-mail: xzlai@outlook.com, wanyili@gdei.edu.cn).
}
\thanks{T. Wu, K. Zhi and M. Elkashlan are with the School of Electronic Engineering and Computer Science at Queen
Mary University of London, London E1 4NS, U.K. (Email:\{tuo.wu,k.zhi, maged.elkashlan\}@qmul.ac.uk).} \thanks{C. Pan is with the National Mobile Communications Research Laboratory, Southeast University, Nanjing 210096, China. (e-mail: cpan@seu.edu.cn).}
}
\markboth{}
{Lai \MakeLowercase{\textit{et al.}}: }

\maketitle

\thispagestyle{empty}

\begin{abstract}
Fluid Antenna System (FAS) is recognized as a promising technology for enhancing communication performance. In this context, we explored the potential of FAS-assisted wireless powered communication systems. Specifically, the transmitter, equipped with FAS, harvests the radio frequency (RF) signal from a power beacon and utilizes the harvested energy for data transmission to the receiver. To evaluate the performance of the considered systems, we derive both the analytical and asymptotic expressions of the outage probability. Simulation results indicate that the diversity of the considered network closely aligns with the number of ports. Besides, it is also revealed that the port selection criteria based solely on single-hop configurations yield a diversity order of only one.

\end{abstract}
\begin{IEEEkeywords}
Fluid antenna system (FAS), wireless powered communications, outage probability.
\end{IEEEkeywords}

\section{Introduction}
Fluid antenna systems (FASs) have emerged as a promising technique for advancing future sixth generation (6G) communications \cite{Wong-frontiers22,MFAS23}. FASs leverage the unique capability of switching the antenna element to an optimal position, referred to as ``ports", with the help of either liquid metal \cite{Huang21, ZJ} or pixel-based antennas  \cite{Rodrigo14, JYao}, and therefore significantly improve the system performance.

Recognizing the significant potential of FAS, the foundational work by Wong \emph{et al.} pioneered analysis in FAS-enabled communication networks, developing the probability density function (PDF) and cumulative distribution function (CDF) for the signal-to-noise ratio (SNR) in point-to-point systems \cite{FAS20,FAS21,Lai23}. Building on this   work, subsequent studies have expanded the application of FAS into multiple access networks \cite{XuH1,XuH2,XuH3,FAMS,FAMS23,Waqar23}. For instance, \cite{XuH1} delves into channel estimation challenges within a multiuser millimeter-wave (mmWave) time-division duplexing (TDD) system. Following this, Xu \emph{et al.} in \cite{XuH2} explored capacity maximization in mmWave systems through the joint optimization of transmit covariance matrices and user antenna position vectors. Then, Xu \emph{et al.}   progressed to examining the outage probability in two-user fluid antenna multiple access systems, as presented in \cite{XuH3}. Additionally, recent advances have seen the integration of machine learning techniques into FAS, particularly for the optimization of port selection \cite{Chai22}.

Besides, the wireless powered communications (WPC)  \cite{WPC19} have also been regarded as a powerful technology, promising to enhance the functionality and efficiency of battery-powered devices. Specifically, the power beacons are employed to charge the battery of the Internet-of-Things (IoT) devices with radio frequency (RF) signal, thereby supporting  the signal transmission of the IoT devices which are placed in hazard environment or inconvenient positions \cite{WPC23}.

Utilizing FAS at mobile devices even with limited space offers a great spatial diversity gain, which matches perfectly with the principles of WPC systems. Specifically, the spatial diversity gain from FAS can significantly enhance the capability of  IoT  devices to utilize charging power more efficiently for data transmission.  This strategy not only boosts efficiency but also promotes more sustainable ways of communication. However, the integration of FAS into WPC systems remains unexplored. To fill this research gap, this paper delves into exploring the potential of the FAS-assisted WPC systems.   To evaluate the impact of network parameters on performance, we have derived both closed-form and asymptotic expressions for outage probability.  Additionally, we present several simulation results, including comparisons of FAS port selection strategies in single hop scenarios.

Specifically, our contributions are summarized as follows:
\begin{itemize}
\item First, we consider a wireless powered system wherein the transmitter is  equipped with a fluid antenna, which can switch the antenna fluid to the optimal port, out of the total $M$ ports. Besides, the transmitter  receives the wireless power from a power beacon, then uses this energy to transmit signal to the receiver at a data rate $R$.
\item We theoretically derive  the analytical and asymptotic expressions of outage probability. Besides, the asymptotic analysis of the outage probability demonstrates that the diversity order of the system under consideration is approximately $M$. Consequently, larger values of $M$ can significantly improve the performance of the FAS-assisted WPC systems.
\item The simulation results substantiate the effectiveness of the proposed analytical approach, thereby confirming and validating our insights and discussions.
\end{itemize}

\section{System Model}
In this study, we examine a wireless powered system wherein the transmitter, receiving wireless power from a power beacon, utilizes this energy to transmit information to the receiver at a data rate $R$. To enhance the performance, the transmitter is equipped with a fluid antenna, which can flexibly adjust the antenna fluid to the optimal port, out of the total $M$ switchable ports. In the following, we will detail the procedure of communications and introduce the FAS channel model, respectively.

\subsection{Communications Model}
To clarify, we define the channel parameters between the power beacon and the $m$-th port of the transmitter as $h_{1,m}\sim\mathcal{CN}(0,\psi_1)$, and those between the $m$-th port of the transmitter and the receiver as $h_{2,m}\sim\mathcal{CN}(0,\psi_2)$. Here, $m\in\mathcal{M}=\{1,2,\cdots,M\}$, $\psi_1$ represents  the average channel gain from the power beacon to the transmitter, and $\psi_2$  denotes the average channel gain from the transmitter to the receiver. Assuming that the $m$-th port is selected for both wireless power harvesting and subsequent data transmission, the power signal received at the transmitter is given as
\begin{align}\label{a1}
g_1=h_{1,m}\sqrt{P_S}b+v_1,
\end{align}
where $P_S$ denotes the transmit power of the power beacon, $b\sim \mathcal{CN}(0,1)$ represents the wireless power signal, and $v_1\sim\mathcal{CN}(0,\sigma^2)$  denotes  the zero-mean complex Gaussian noise. Therefore, the total harvested energy at the transmitter is $|h_{1,m}|^2P_ST_1$, where $T_1$ is the time duration of the first hop.

Then, the transmitter   utilizes the harvested wireless energy to transmit data $s\sim\mathcal{CN}(0,1)$. Consequently, the signal received at the receiver is written as
\begin{align}\label{a2}
g_2=h_{2,m}|h_{1,m}|\sqrt{P_S \rho \theta }s+v_2,
\end{align}
where $v_2\sim\mathcal{CN}(0,\sigma^2)$ represents the zero-mean complex Gaussian noise, $\rho$ denotes the energy conversion efficiency, $\theta=T_1/T_2$, and $T_2$ is the time duration of the second hop.

Drawing on the analysis from Eq. \eqref{a1} and Eq. \eqref{a2},  the received SNR at the receiver utilizing the $m$-th port can be formulated as
\begin{align}\label{a3}
\Gamma_m=\frac{P_S\rho\theta}{\sigma^2}\gamma_{1,m}\gamma_{2,m},
\end{align}
where  $\gamma_{i,m}=|h_{i,m}|^2$ for $i\in\{1,2\}$ represent the channel gains of the first and second hops, respectively.

\subsection{FAS Channel Model}
In this subsection, we present the details of the channel model of FAS under the Rayleigh fading environments \cite{Lai23}, which accounts for the combined impact of port correlation. It is important to note that while some more intricate models are available, such as those described in \cite{Khammassi23}, they often yield results that are impractical for widespread applications due to their complexity.   Building on the Rayleigh fading channel model, the expression for $h_{i,m}$ is expressed as
\begin{align}\label{a4}
h_{i,m}=\sqrt{\mu}h_{i,0}+\sqrt{1-\mu} e_{i,m},
\end{align}
where  $e_{i,m}\sim\mathcal{CN}(0,\psi_i)$ for $m\in\mathcal{M}$ are independently and identically distributed (i.i.d.) random variables (RVs). Here,  $\psi_i$ denotes the average channel gain both from the power beacon to the transmitter and from the transmitter to the receiver.
Moreover, $h_{1,0}\sim\mathcal{CN}(0,\psi_1)$   and $h_{2,0}\sim\mathcal{CN}(0,\psi_2)$ serve as virtual reference channels for these respective links. Consequently,  the PDFs of $\gamma_{i,0}=|h_{i,0}|^2$ for $i \in\{1,2\}$ can be expressed as
\begin{align}\label{aa5}
f_{\gamma_{i,0}}(x)=\frac{1}{\psi_i}e^{-\frac{x}{\psi_i}}.
\end{align}
Additionally, $\mu$ represnts the correlation factor, which is given by \cite{FAS22}
\begin{align}\label{a6}
\mu=&\sqrt{2}\sqrt{{}_1F_2\Big(\frac{1}{2};1;\frac{3}{2};-\pi^2W^2\Big)-\frac{J_1(2\pi W)}{2\pi W}},
\end{align}
where
 ${}_a F_b$ denotes the generalized hypergeometric function and $J_1(\cdot)$ is the first kind  Bessel function with order 1.

Given a fixed channel parameter $h_{i,0}$, and in accordance with $\gamma_{i,0}$, the associated channel gain for $h_{i,m}$, expressed as $\gamma_{i,m}=|h_{i,m}|^2$, follows a non-central chi-square distribution. Consequently, the conditional PDF  can be expressed as
\begin{align}\label{a7}
f_{\gamma_{1,m} | \gamma_{1,0}=x_0}(x)=&\omega_1
e^{-\omega_1(x+\mu x_{0})}
I_{0}\big(2\omega_1\sqrt{\mu x_{0} x}\big),\\ \label{a8}
f_{\gamma_{2,m} | \gamma_{2,0}=y_0}(y)=&\omega_2
e^{-\omega_2(y+\mu y_{0})}
I_{0}\big(2\omega_2\sqrt{\mu y_{0} y}\big),
\end{align}
where  $\omega_i=\big(\psi_i(1-\mu)\big)^{-1}$. Besides, $I_{0}(u)$ is the modified Bessel function of the first kind with order $0$, which  can be expressed in series representation as \cite{book}
\begin{align}\label{a9}
I_0(z)=\sum_{k=0}^{\infty}\frac{z^{2k}}{2^{2k} k!\Gamma(k+1)}.
\end{align}
Combining  \eqref{a7}--\eqref{a9}, we further derive $f_{\gamma_{1,m} | \gamma_{1,0}=x_0}(x)$ and $f_{\gamma_{2,m} | \gamma_{2,0}=y_0}(y)$ as
\begin{align}\label{a10}
f_{\gamma_{1,m} | \gamma_{1,0}=x_0}(x)=&
\sum_{k=0}^{\infty}a_k x_0^k e^{-\omega_1\mu x_{0}} x^k e^{-\omega_1 x},\\
f_{\gamma_{2,m} | \gamma_{2,0}=y_0}(y)=&
\sum_{n=0}^{\infty}b_n y_0^n e^{-\omega_2\mu y_{0}} y^n e^{-\omega_2 y},
\end{align}
where
\begin{align}\label{a11}
a_k=&\frac{\omega_1^{2k+1}\mu^{k}}{(k!)^2},\notag\\
b_n=&\frac{\omega_2^{2n+1}\mu^{n}}{(n!)^2}.
\end{align}

In communication networks equipped with FAS, the transmitter possesses the capability to switch the antenna fluid to the optimal port, thereby having additional degree-of-freedom to improve the system performance. To maximize the received SNR at the receiver, the selection of the port for transmission is governed by the following criterion:
\begin{align}\label{a12}
m^*=\arg\max_{m\in\mathcal{M}}(\gamma_{1,m}\gamma_{2,m}).
\end{align}
Hence, the received SNR at the receiver  is given by
\begin{align}\label{a13}
\Gamma_{m^*}=\frac{P_S\rho\theta}{\sigma^2}\max_{m\in\mathcal{M}}(\gamma_{1,m}\gamma_{2,m}).
\end{align}

\section{Outage Analysis}
In this section, we derive the closed-form expression of the outage probability. Subsequently, the  asymptotic expression of the outage probability is derived. These derivations offer valuable insights for the practical application of the FAS-enabled WPC system.

\subsection{Closed-Form Outage Probability}
The outage of communications happens when the received SNR falls below the given threshold $\Gamma_{\rm{th}}$, i.e., $\Gamma_{\rm{th}}=2^R-1$. Consequently, the outage probability is given by
\begin{align}\label{b1}
P_{\mathrm{out}}=\Pr\left(\max_{m\in\mathcal{M}}(\gamma_{1,m}\gamma_{2,m})\leq z \right),
\end{align}
where $z=\frac{\Gamma_{\rm{th}}\sigma^2}{P_S\rho\theta}$.

To derive the analytical expression for the outage probability $P_{\mathrm{out}}$, we reformulate $P_{\mathrm{out}}$ as
\begin{align}\label{b2}
&P_{\mathrm{out}}\notag\\
&=\mathbf{E}_{\gamma_{1,0},\gamma_{2,0}}\left[F(z\big| \gamma_{1,0}, \gamma_{2,0})\right]\notag\\
&=\int_{0}^{\infty}\int_{0}^{\infty}F(z\big| x_0, y_0)f_{\gamma_{1,0}}(x_0)f_{\gamma_{2,0}}(y_0)dx_0 dy_0,
\end{align}
where $\mathbf{E}_{X}\left[\cdot \right]$ denotes the expectation over RV $X$. Besides,
$F(z\big| x_0, y_0)$ represents the conditional CDF with given $\gamma_{1,0}=x_0$ and $\gamma_{2,0}=y_0$, which is defined as
\begin{align}\label{b3}
&F(z\big| x_0, y_0)\notag\\
&=
\Pr\big(\max_{m\in\mathcal{M}}(\gamma_{1,m}\gamma_{2,m})\leq z\big| \gamma_{1,0}=x_0, \gamma_{2,0}=y_0 \big).
\end{align}
Expanding on this, and considering that $\gamma_{1,m}\gamma_{2,m}$ for $m\in\cal{M}$ are independent with each other, the conditional CDF $F(z\big| x_0, y_0)$, when conditioned on $\gamma_{1,0}=x_0, \gamma_{2,0}=y_0$, can be further rewritten as
\begin{align}\label{b4}
F(z\big| x_0, y_0)=\big(\Phi(z\big| x_0,y_0)\big)^M,
\end{align}
where $\Phi(z\big| x_0,y_0)$ is defined as
\begin{align}\label{b5}
&\Phi(z\big| x_0,y_0)\notag\\
&=\Pr\left(\gamma_{1,m}\gamma_{2,m}\leq z \big| \gamma_{1,0}=x_0, \gamma_{2,0}=y_0 \right)\notag\\
&=\int_{0}^{\infty}\int_{0}^{z/y}f_{\gamma_{1,m} | \gamma_{1,0}=x_0}(x)f_{\gamma_{2,m} | \gamma_{2,0}=y_0}(y)dx dy
\end{align}
Building upon this, further mathematical manipulations are applied to calculate this function. The detailed derivation leads to
\begin{align}\label{b6}
&\Phi(z\big| x_0,y_0)\notag\\
&=\sum_{k=0}^{\infty}\sum_{n=0}^{\infty}a_k b_n x_0^k e^{-\omega_1\mu x_0}  y_0^n e^{-\omega_2\mu y_0} \notag\\
&\quad \times\int_{0}^{\infty}\int_{0}^{\frac{z\omega_1}{y}}
x^ke^{-\omega_1 x}y^ne^{-\omega_2y}dxdy
\notag\\
&=
\sum_{k=0}^{\infty}\sum_{n=0}^{\infty}\frac{a_k b_n x_0^k e^{-\omega_1\mu x_0}   y_0^n e^{-\omega_2\mu y_0} }{\omega_1^{k+1}}\notag\\
&\quad\times\int_{0}^{\infty}\gamma\Big(k+1,\frac{z\omega_1}{y}\Big)y^ne^{-\omega_2y}dy
\end{align}
where $\gamma(\alpha,x)$ is the upper Gamma function, which can be expressed as
\begin{align}\label{b7}
\gamma(\alpha,x)
&=\int_0^x e^{-t}t^{\alpha-1}dt\notag\\
&=(\alpha-1)!\bigg(1-e^{x}\sum_{d=0}^{\alpha-1}\frac{x^d}{d!}\bigg).
\end{align}
Applying \eqref{b7} in \eqref{b6}, $\Phi(z\big| x_0,y_0)$ can be expressed in an series representation as
\begin{align}\label{b8}
&\Phi(z\big| x_0,y_0)\notag\\
&=\sum_{k=0}^{\infty}\sum_{n=0}^{\infty}\frac{a_k b_n x_0^k e^{-\omega_1\mu x_0} k!  y_0^n e^{-\omega_2\mu y_0} }{\omega_1^{k+1}}\notag\\
&\quad\times\bigg(
\int_0^{\infty}y^n e^{-\omega_2y}dy-\sum_{l=0}^k\int_0^{\infty}e^{-\frac{z\omega_1}{y}-\omega_2y}\frac{(z\omega_1)^l}{l!y^l }dy
\bigg)\notag\\
&=\sum_{k=0}^{\infty}\sum_{n=0}^{\infty}\frac{a_k b_n x_0^k e^{-\omega_1\mu x_0} k!  y_0^n e^{-\omega_2\mu y_0} }{\omega_1^{k+1}\omega_2^{n+1}}\notag\\
&\quad\times\big(n!-
\sum_{l=0}^{k}\frac{2(z\omega_1\omega_2)^{\frac{n+l+1}{2}}}{l!}K_{n-l+1}(2\sqrt{z\omega_1\omega_2})
\big),
\end{align}
where $K_{\nu}(\cdot)$ denotes the second kind modified Bessel function with order $\nu$.
The final expression  in \eqref{b8} is derived using the result from \cite[3.471.9]{book}, that is
\begin{align}
\int_{o}^{\infty}x^{a-1}e^{-\frac{b}{x}-cx}dx=2\Big(\frac{b}{c}\Big)^{\frac{a}{2}}K_{a}(2\sqrt{bc}).
\end{align}

Then, by substituting \eqref{b8} into \eqref{b4}, we can achieve
\begin{align}\label{b10}
&F(z\big| x_0, y_0)\notag\\
&=\sum_{k_m, m\in\mathcal{M}}^{\infty}\sum_{n_m, n\in\mathcal{M}}^{\infty} D_{k,n}
x_0^{\epsilon_k} y_0^{\lambda_n}\notag\\
&\quad\times
e^{-M\omega_1\mu x_0}e^{-M\omega_2\mu y_0},
\end{align}
where
\begin{align}\label{b11}
&D_{k,n}\notag\\
&=\prod_{m=1}^M  \big(n_m!-\sum_{l=0}^{k_m}\frac{2(z\omega_1\omega_2)^{\frac{n_m+l+1}{2}}}{l!}K_{n_m-l+1}(2\sqrt{z\omega_1\omega_2})\big)\notag\\
&\quad\times\frac{\omega_1^{k_m} \omega_2^{n_m} \mu^{k_m+n_m}}{k_m!(n_m!)^2}, \notag\\
&\epsilon_k=\sum_{m=1}^M k_m, \notag\\
 & \lambda_n=\sum_{m=1}^M n_m.
\end{align}
To derive the analytical expression of the outage probability $P_{\rm{out}}$, we substitute the expressions for $F(z\big| x_0, y_0)$ from  \eqref{b10} and $f_{\gamma_{i,0}}(x)$ from  \eqref{aa5} into \eqref{b2}. This process yields
\begin{align}\label{b12}
&P_{\rm{out}}\notag\\
&=\sum_{k_m, m\in\mathcal{M}}^{\infty}\sum_{n_m, n\in\mathcal{M}}^{\infty} \frac{D_{k,n}}{\psi_1\psi_2}
\notag\\
&\quad\times
\int_0^{\infty}x_0^{\epsilon_k} e^{-M\omega_1\mu \frac{x_0}{\psi_1}}dx_0\times \int_0^{\infty} y_0^{\lambda_n} e^{-M\omega_2\mu \frac{y_0}{\psi_2}}dy_0\notag\\
&=\sum_{k_m, m\in\mathcal{M}}^{\infty}\sum_{n_m, n\in\mathcal{M}}^{\infty} D_{k,n}
\frac{\epsilon_k!\lambda_n!}{q_1^{\epsilon_k+1}q_2^{\lambda_n+1}\psi_1\psi_2}\notag\\
&\approx \sum_{t=0}^{U_k\times M}\sum_{u=0}^{U_n\times M}\frac{t!u!}{q_1^{t+1}q_2^{u+1}\psi_1\psi_2}
\sum_{\mathcal{P}_{k}(t,U_k)}\
\sum_{\mathcal{P}_{n}(u,U_n)}
D_{k,n},
\end{align}
where $\mathcal{P}_{k}(t,U_k)$ denotes the set including all compositions of $k_m\in \cal{M}$, and in each composition $\sum_{m=1}^{M}k_m=t$ and  $\forall k_m\leq U_k$ hold. $\mathcal{P}_{n}(u,U_n)$ denotes the set including all compositions of $n_m\in \cal{M}$, and in each composition $\sum_{m=1}^{M}n_m=u$ and  $\forall n_m\leq U_n$ hold.
In addition, the approximation in \eqref{b12} involves truncation in practice, with truncation numbers $U_1$ and $U_2$.  The corresponding parameters are defined as
\begin{align}\label{b13}
q_1=&M\omega_1\mu+\frac{1}{\psi_1},\notag\\
 q_2=&M\omega_2\mu+\frac{1}{\psi_2}.
\end{align}
\subsection{Asymptotic Analysis}
To more effectively assess the outage performance of the system under consideration, a theorem is proposed as following.
\begin{theorem}
In the high SNR region,  particularly for small values of $z$, the asymptotic expression of outage probability is
\begin{align}\label{b14}
P{\rm{out}}\approx \chi\big(\beta\ln(\beta)\big)^M,
\end{align}
where
\begin{align}\label{b15}
&\chi=\Big(\frac{1-\mu}{M\mu+1-\mu}\Big)^2,\notag\\ &\beta=\frac{\Gamma_{\rm{th}}\sigma^2}{P_S\rho\phi_1\phi_2(1-\mu)^2}.
\end{align}
\end{theorem}

\emph{Proof 1:} By setting $U_1$ and $U_2$ to 0,  and then applying the approximations $e^{-x}=1-x$ and $K_0(x)\approx -\ln(x)$ for tiny value of $|x|$ \cite{book2}, we can easily obtain the asymptotic expression.
$\hfill\blacksquare$

Accordingly, the following remark can be made concerning the proposed FAS-assisted WPC systems.
\begin{remark}
It is demonstrated that the diversity order of the system under consideration is approximately $M$. Consequently, larger values of $M$ can significantly enhance the performance of the FAS-assisted  WPC systems. Furthermore, as evident from equation \eqref{b11}, the outage probability $P{\rm{out}}$ inversely correlates with the parameter $\mu$. Therefore, with an increased value of $W$,  the efficiency of the FAS-assisted WPC systems can be further improved.
\end{remark}

\section{Numerical Results}
In this section, simulation results are presented to validate the proposed analysis. For the simulations, the parameters are set as follows: $\rho = 0.5$, $\theta = 1$. Unless otherwise specified, the data rate $R$ is fixed at $1$ bit/s/Hz. The average channel gains for each hop are normalized to unity, implying that $\psi_1 = \psi_2 = 1$.

\begin{figure}[t]
\centering
\includegraphics[width=1.1\linewidth]{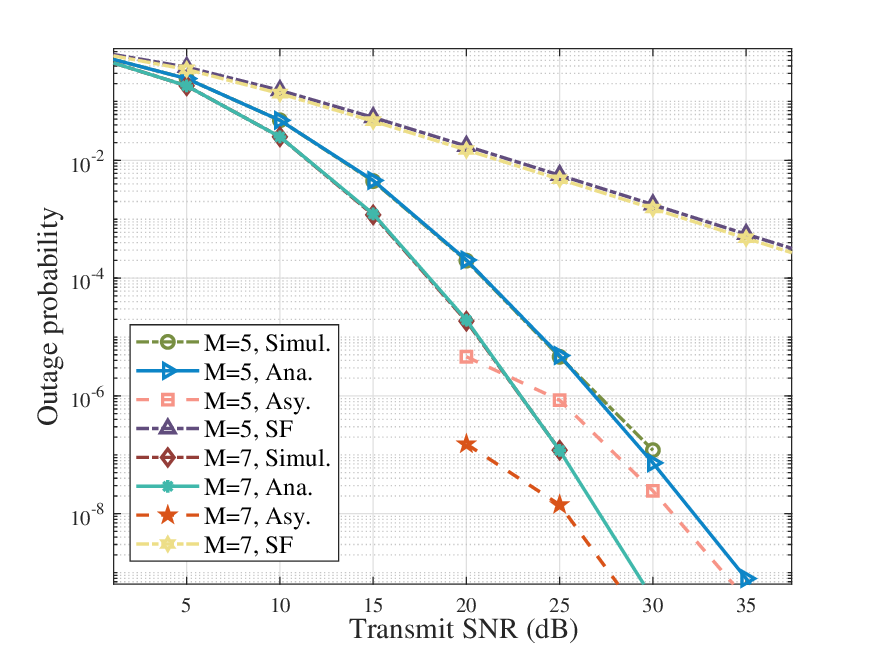}
\caption{Outage probability versus  transmit SNR $P_S/\sigma^2$.}\label{fig1}
\end{figure}

Fig.~\ref{fig1} illustrates the performance of the considered systems, across various transmit SNR, denoted as $P_S/\sigma^2$. In these simulations, $M$ is set to either $5$ or $7$, and $W = 0.6$ is considered. For comparative purposes, the results for a single-hop FAS network, labeled `SF' in the legend, are also included. In the single-hop scenario, the FAS at the transmitter switches the port based solely on either the first or the second hop, which means that the received SNR at the receiver is given by
\begin{align}
\Gamma_{\tilde{m}}=\frac{P_S\rho\theta}{\sigma^2}\max_{m\in\mathcal{M}}(\gamma_{1,m})\gamma_{2,m},
\end{align}
or
\begin{align}
\Gamma_{\tilde{m}}= \frac{P_S\rho\theta}{\sigma^2}\gamma_{1,m}\max_{m\in\mathcal{M}}(\gamma_{2,m}).
\end{align}
As observed, the analytical results align perfectly with the simulation results, and the asymptotic results converge with the simulation outcomes at higher transmit SNR levels, which validate the correctness of the derivations.

 Moreover, Fig.~\ref{fig1} not only confirms the expected trend of decreasing outage probability with increasing transmit SNR or the number of ports $M$ but also reveals the slight difference in the performance of the single-hop FAS-assisted systems. Notably, while an increase in $M$ generally enhances system performance, the improvement becomes marginally less significant in single-hop FAS-assisted systems as $M$ grows larger. This observation suggests diminishing returns on the performance gains from increasing the number of ports beyond a certain point in single-hop scenarios. Such insights could be pivotal for optimizing resource allocation and system design in FAS-assisted systems.

\begin{figure}[t]
\centering
\includegraphics[width=1.1\linewidth]{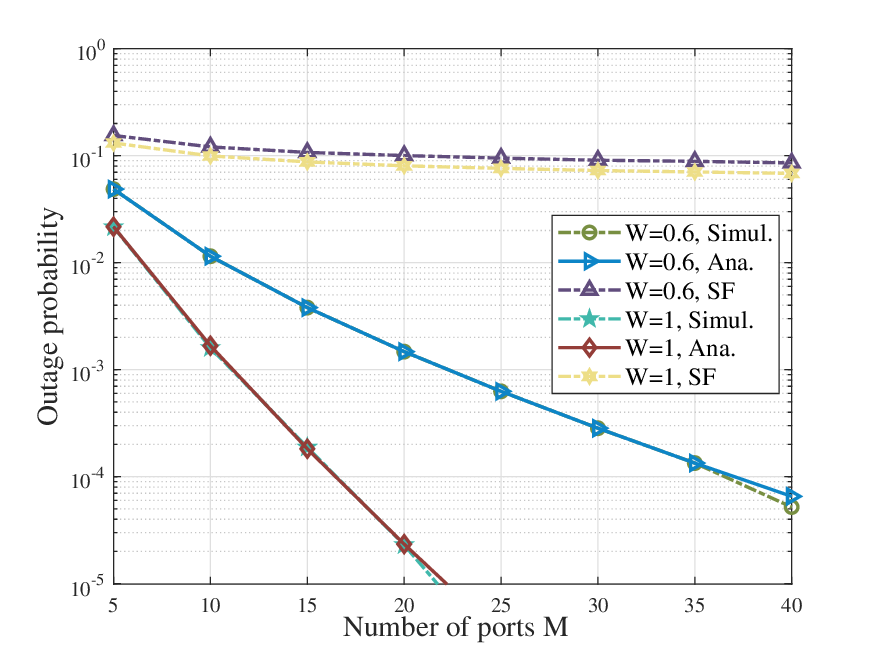}
\caption{Outage probability versus different number of ports $M$.}\label{fig2}
\end{figure}

Fig.~\ref{fig2} demonstrates the change in outage probability across different port numbers, $M$, ranging from 5 to 40, under a constant transmit SNR of  $P_S/\sigma^2 = 10$ dB. This figure distinctly shows that within the proposed FAS network, the outage probability diminishes as the number of ports, $M$,  increases. Notably, this trend is more pronounced with larger values of $W$, aligning with the predictions made in Theorem 1. These results underscore the importance of port number in enhancing network performance, particularly in scenarios with higher values of $W$.

\begin{figure}[t]
\centering
\includegraphics[width=1.1\linewidth]{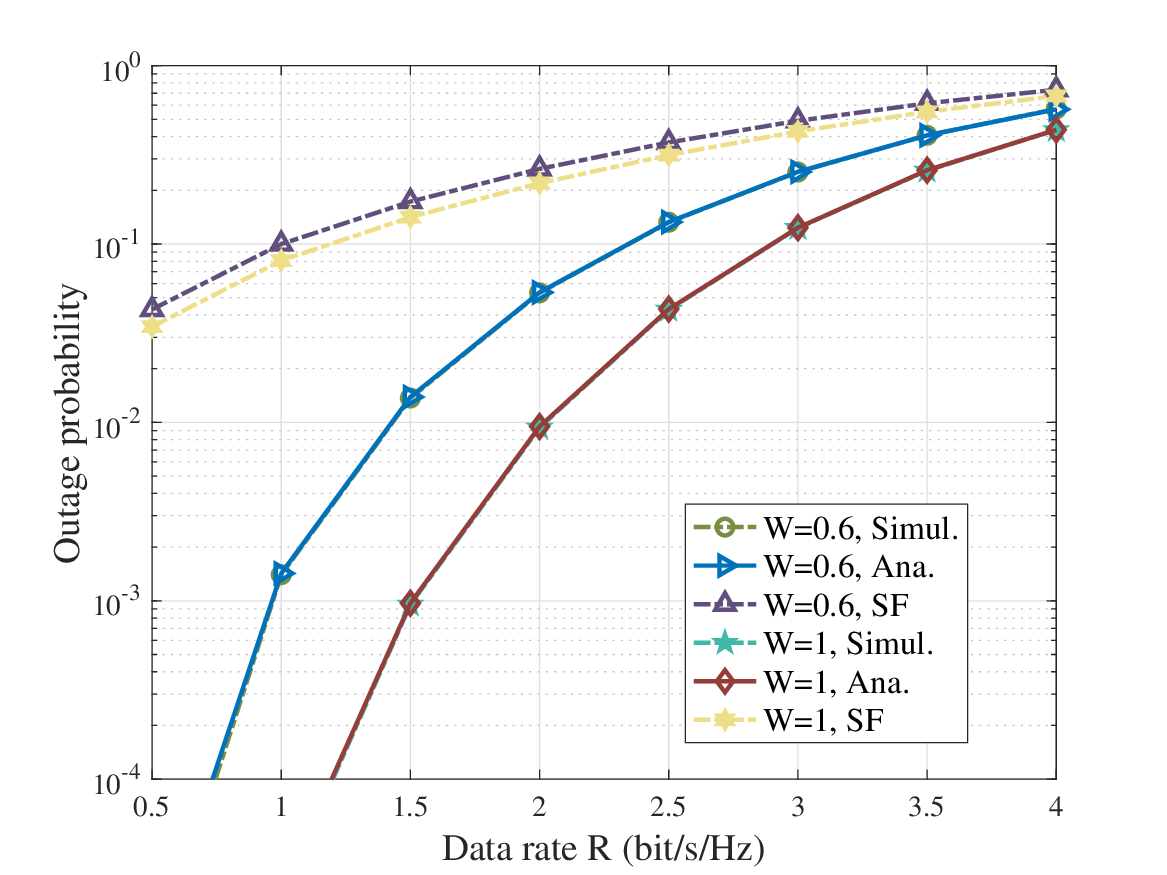}
\caption{Outage probability versus different data rate $R$.}\label{fig3}
\end{figure}

Fig.~\ref{fig3} showcases the variation in outage probability with different data rate values, i.e., $R$, within a range from 0.5 to 4 bit/s/Hz. This is observed in a transmission environment with an SNR of $P_S/\sigma^2 = 10$ dB and $M=20$. As depicted in Fig.~\ref{fig3}, there is a notable increase in the outage probability in the proposed FAS-assisted communication systems as the data rate $R$ escalates. This increase is particularly significant for larger values of $W$, affirming the theoretical predictions presented in Theorem 1. These results demonstrate the impact of varying data rates on network performance, especially in scenarios with higher values of $W$.

\section{Conclusions}
In this work, we proposed a performance analysis of the FAS-assisted WPC  systems. We derived both exact and asymptotic expressions for the outage probability of the proposed system with Rayleigh fading channels. Our analysis revealed that the diversity order of the system is equal to the number of ports. Numerical results were provided to demonstrate the correctness of our analysis.



\begin{thebibliography}{1}

\bibitem{Wong-frontiers22}
K. K. Wong, K. F. Tong, Y. Shen, Y. Chen, and Y. Zhang, ``Bruce Lee-inspired fluid antenna system: Six research topics and the potentials for 6G,'' {\em Frontiers Commun. Netw., section Wireless Commun.}, 3:853416, Mar. 2022.

\bibitem{MFAS23} K.-K. Wong, K.-F. Tong, and C.-B. Chae, ``Fluid antenna system-part II: Research opportunities," \emph{IEEE Commun. Lett.}, vol. 27, no. 8, pp. 1924--1928, Aug. 2023.


\bibitem{Huang21}Y. Huang, L. Xing, C. Song, S. Wang, and F. Elhouni, ``Liquid antennas: Past, present and future," \emph{IEEE Open J. Antennas Propag.}, vol. 2, pp. 473--487, 2021.

\bibitem{Rodrigo14}D. Rodrigo, B. A. Cetiner, and L. Jofre, ``Frequency, radiation pattern and polarization reconfigurable antenna using a parasitic pixel layer," \emph{IEEE Trans. Antennas Propag.}, vol. 62, no. 6, pp. 3422--3427, Jun. 2014.

\bibitem{ZJ} J. Zheng, et. al. ``FAS-assisted NOMA Short-Packet Communication Systems."  \emph{arXiv preprint}, arXiv:2310.14251.
    
\bibitem{JYao} J. Yao, et. al. ``Proactive Monitoring via Jamming in Fluid Antenna Systems."  \emph{arXiv preprint}, arXiv:2310.07550.

\bibitem{FAS20}K.-K. Wong, A. Shojaeifard, K.-F. Tong and Y. Zhang, ``Performance limits of fluid antenna systems," \emph{IEEE Commun. Letters}, vol. 24, no. 11, pp. 2469--2472, Nov. 2020.

\bibitem{FAS21}K.-K. Wong, A. Shojaeifard, K.-F. Tong, and Y. Zhang, ``Fluid antenna systems," \emph{IEEE Trans. Wireless Commun.}, vol. 20, no. 3, pp. 1950--1962, March 2021.


\bibitem{Lai23} X. Lai, T. Wu, J. Yao, C. Pan, M. Elkashlan, and K. -K. Wong, ``On performance of fluid antenna system using maximum ratio combining," \emph{IEEE Commun. Lett.}, doi: 10.1109/LCOMM.2023.3348028.

\bibitem{XuH1} H. Xu, et. al. ``Channel estimation for FAS-assisted multiuser mmWave systems," \emph{IEEE Commun. Lett.}, 2023.

\bibitem{XuH2} H. Xu, et. al. ``Capacity maximization for FAS-assisted multiple access channels," \emph{arXiv preprint}, arXiv:2311.11037.

\bibitem{XuH3} H. Xu, et. al.  ``On outage probability for two-user fluid antenna multiple access," \emph{ IEEE International Conference on Communications}, IEEE, 2023.

\bibitem{FAMS}K.-K. Wong and K.-F. Tong, ``Fluid antenna multiple access," \emph{IEEE Trans. Wireless Commun.}, vol. 21, no. 7, pp. 4801-C4815, Jul. 2022.


\bibitem{FAMS23}K.-K. Wong, D. Morales-Jimenez, K.-F. Tong, and C.-B. Chae, ``Slow fluid antenna multiple access," \emph{ IEEE Trans. Commun.}, vol. 71, no. 5, pp. 2831--2846, May 2023.

\bibitem{Waqar23}N. Waqar, K.-K. Wong, K.-F. Tong, A. Sharples, and Y. Zhang, ``Deep learning enabled slow fluid antenna multiple access," \emph{IEEE Commun. Lett.}, vol. 27, no. 3, pp. 861--865, March 2023.

\bibitem{Chai22}Z. Chai, K.-K. Wong, K.-F. Tong, Y. Chen, and Y. Zhang, ``Port selection for fluid antenna systems," \emph{IEEE Commun. Lett.}, vol. 26, no. 5, pp. 1180--1184, May 2022.



\bibitem{Khammassi23}M. Khammassi, A. Kammoun and M. -S. Alouini, ``A new analytical approximation of the fluid antenna system channel,'' {\em IEEE Trans. Wireless Commun.}, vol. 22, no. 12, pp. 8843--8858, Dec. 2023.


\bibitem{WPC19}K. Liang, L. Zhao, G. Zheng, and H. -H. Chen, ``Non-uniform deployment of power beacons in wireless powered communication networks," \emph{IEEE Trans. Wireless Commun.}, vol. 18, no. 3, pp. 1887-1899, March 2019.

\bibitem{WPC23}D. M. Mughal, D. Munir, and M. Y. Chung, ``Outage analysis of IRS-assisted RF powered networks for energy-constrained IoT devices," \emph{IEEE Trans. Wireless Commun.}, vol. 22, no. 11, pp. 7805--7818, Nov. 2023.

\bibitem{Xie23}K. Xie, G. Cai, G. Kaddoum, and J. He, ``Performance analysis and resource allocation of STAR-RIS-aided wireless-powered NOMA system," \emph{IEEE Trans. Wireless Commun.}, vol. 71, no. 10, pp. 5740--5755, Oct. 2023.

\bibitem{FAS22}K.-K. Wong, K. F. Tong, Y. Chen, and Y. Zhang, ``Closed-form expressions for spatial correlation parameters for performance analysis of fluid antenna systems," \emph{IET Electron. Lett.}, vol. 58, no. 11, pp. 454--457, May 2022.

\bibitem{book}I. S. Gradshteyn and I. M. Ryzhik, \emph{Table of Integrals, Series, and Products, 7th ed}. San Diego, CA: Academic, 2007.

\bibitem{book2} M. Abramowitz, I. A. Stegun, \emph{Handbook of mathematical functions: With formulas, graphs, and mathematical tables, 1st ed}. Nassau County, NY: Dover, 1965.
\end{thebibliography}
\end{document}